\documentclass[aps,prb,twocolumn,floatfix,superscriptaddress,amssymb,amsmath]{revtex4}

\usepackage{graphicx}

\begin{document}

\title{Decoherence of Rabi oscillations in a single quantum dot}

\author{J.~M.~Villas-B\^{o}as}
\affiliation{Department of Physics and Astronomy, and Nanoscale
and Quantum Phenomena Institute, \\Ohio University, Athens, Ohio
45701-2979} \affiliation{Departamento de F\'{\i}sica, Universidade
Federal de S\~{a}o Carlos, 13565-905, S\~{a}o Carlos, S\~{a}o
Paulo, Brazil}

\author{Sergio E.~Ulloa}
\affiliation{Department of Physics and Astronomy, and Nanoscale
and Quantum Phenomena Institute, \\Ohio University, Athens, Ohio
45701-2979}

\author{A.~O.~Govorov}
\affiliation{Department of Physics and Astronomy, and Nanoscale
and Quantum Phenomena Institute, \\Ohio University, Athens, Ohio
45701-2979}


\begin{abstract}
We develop a realistic model of Rabi oscillations in a quantum-dot
photodiode. Based in a multi-exciton density matrix formulation we
show that for short pulses the two-level models fails and higher
levels should be taken into account. This affects some of the
experimental conclusions, such as the inferred efficiency of the
state rotation (population inversion) and the deduced value of the
dipole interaction. We also show that the damping observed cannot
be explained using \emph{constant} rates with fixed pulse
duration. We demonstrate that the damping observed is in fact
induced by an off-resonant excitation to or from the continuum of
wetting layer states. Our model describes the nonlinear
decoherence behavior observed in recent experiments.
\end{abstract}

\maketitle

Coherent manipulation of a quantum state is one of the tasks
required in quantum computation and information processing.
Semiconductor quantum dots (QDs) have been shown to be excellent
candidates for the physical implementation of these objectives.
Rabi oscillations -- temporal coherent oscillations of the
population and its inversion in a two-level system driven by a
strong resonant field -- have been successfully demonstrated by
different groups using excitons in single QDs with different
clever probing techniques,\cite{Stievater01,Kamada01,Htoon02,%
Zrenner02,Borri02,Besombes03,Muller04,Wang04} as well as utilizing
the biexciton population.\cite{Li03} Zrenner \textit{et
al.}\cite{Zrenner02} have developed a single self-assembled QD
photodiode in which the population inversion is probed by the
photocurrent signal induced by a strong and carefully tuned
optical pulse. In their device the pulse generates an
electron-hole pair in the QD and with the help of an external gate
voltage, the electron and hole tunnel out into nearby contacts.
This process generates a photocurrent signal that is a weakly
disturbing probe of the coherent state of the system. Their
results show Rabi oscillations that are damped with increasing
area of the pulse for a fixed pulse duration. However, the
mechanisms that produce this decoherence were unclear.

We present a study of the dynamics of a single self-assembled
quantum dot photodiode in the presence of an optical pulse.  The
electron and hole tunneling processes are introduced via a
microscopic model of the structure which also includes the
electron-hole interaction. The dynamics is described using a density
matrix approach that incorporates dipole coupling to multi-exciton
states and off-resonant excitation to states in the wetting layer
(WL). Our model shows that for short pulses (of the order of a few
ps, as used in experiments\cite{Zrenner02}) the two level system
approximation fails for a $\pi$-pulse (the pulse necessary to invert
the exciton population) and the biexciton state plays an important
role in the dynamics. This results in a frequency shift of the
measured photocurrent oscillations, which can significantly affect
the experimentally deduced value of the transition dipole moment.
The efficiency estimate of the state rotation (population inversion)
for a $\pi$-pulse is also affected as the biexciton contributes to
the photo-signal. We demonstrate that longer pulses minimize the
biexciton contribution. Most importantly, we find that inclusion of
excitations to WL states is essential to understand the decoherence
observed in experiments. The damping due to coupling with the WL not
only explains the observed shape of Rabi oscillations, but also
gives a description of the background signal observed in
Rabi-photodiodes.\cite{Beham03} Although we focus on the QD
photodiode,\cite{Zrenner02} our model is also relevant for
experiments on Rabi oscillations with optical readout.\cite{Wang04}

\begin{figure}[tbp]
\includegraphics*[width=1.0\linewidth]{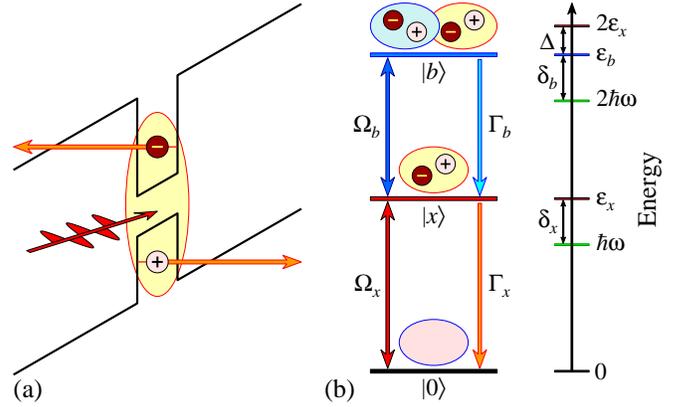}
\caption{Schematic band structure and level configuration of QD
photodiode. (a) An electromagnetic pulse creates an exciton in the
dot, and an applied gate voltage forces the electron and hole to
tunnel out, generating the photocurrent signal. (b) Schematic
representation of processes and levels involved in this system:
$|x\rangle$ and $|b\rangle$ are exciton and biexciton states.}
\label{fig1}
\end{figure}

In Fig.\ \ref{fig1} we show the system and level configuration
taken into account by our model Hamiltonian which can be written
as\cite{Boas04}
\begin{eqnarray}
H_0 &=&\delta_{x}|x\rangle \langle x|+\delta_{b}|b\rangle \langle b|
\nonumber \\
&&-\frac{1}{2}\Bigl[\Omega_x(t)|0\rangle
\langle x| +\Omega_b(t)|x\rangle \langle b| +h.c.\Bigr].
\label{eq1}
\end{eqnarray}
Here, $\delta_{x}=\varepsilon_x-\hbar\omega$ accounts for the
detuning of the exciton with the laser energy $\hbar \omega$,
$\delta_{b}=\varepsilon_b-2\hbar\omega$ is the two photon
biexciton detuning, $\Omega_{x}(t)=\langle 0| \vec{\mu}\cdot
\vec{E}(t)|x\rangle/\hbar$, $\Omega_{b}(t)=\langle x|
\vec{\mu}\cdot \vec{E}(t)|b\rangle/\hbar$, where the electric
dipole moment $\vec{\mu}$ describes the coupling of the excitonic
transition to the radiation field, and $\vec{E}(t)$ is the pulse
amplitude which we assume to have a Gaussian shape, with full
width at half-maximum (FWHM) $t_p$ (pulse length).

\begin{figure}[tbp]
\includegraphics*[width=1.0\linewidth]{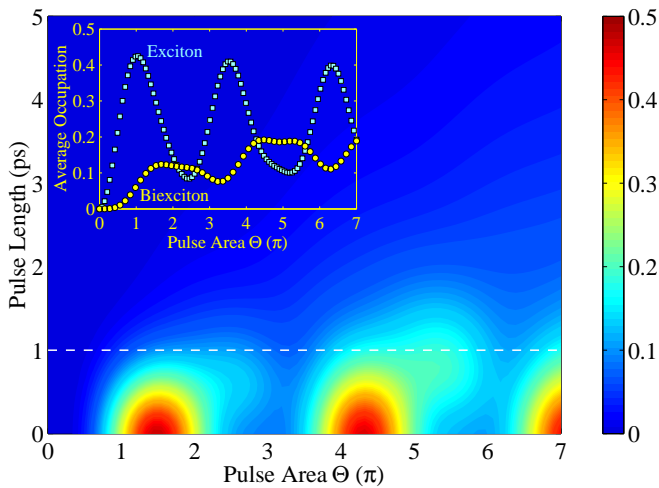}
\caption{Contour plot of the biexciton average occupation as
function of pulse duration $t_p$ and pulse area $\Theta$. Notice
long pulses do not produce biexciton population. Inset compares
exciton and biexciton contribution for a 1 ps pulse.} \label{fig2}
\end{figure}

For a strong pulse resonant with the exciton energy
($\delta_{x}=0$), the biexciton binding energy $\Delta=3$ meV is
relatively small and, as can be seen in Fig.\ \ref{fig2}, cannot
be neglected in the dynamical description of the system. There we
show the average occupation of the biexciton state as a function
of the pulse length $t_p$ and pulse area
$\Theta=\int_{-\infty}^{\infty}\Omega(t)dt$, assuming that the
biexciton has the same dipole moment of the exciton
transition\cite{Chen02} [$\Omega_b(t)=\Omega_x(t)=\Omega(t)$] and
not including decay or dephasing.  We show below that even after
inclusion of these effects the biexciton population is important
for short pulses.

The dynamics of the reduced system is computed using a master
equation in the Lindblad form\cite{Mahler98}
\begin{equation}
\frac{d\rho}{dt} = -\frac{i}{\hbar}[H_0, \rho] + L(\rho),
\label{eq2}
\end{equation}
where the first term on the right yields the unitary evolution of
the quantum system and $L(\rho)$ is the dissipative part of the
evolution assuming the Markovian approximation. We use $\Gamma_i$ to
describe all population decay rates of the level $i$, which we
assume to have two different sources, one due to the spontaneous
decay $\Gamma_i^\text{rec}$ given by the recombination rate (for the
exciton state the spontaneous decay time is known to be
$\tau_\text{rec}=1/\Gamma_x^\text{rec}\simeq 1$ ns for this kind of
QD) and other $\Gamma_i^\text{tun}$ which results when the particles
(making the exciton or biexciton) leave the system by tunneling, so
that $\Gamma_i=\Gamma_i^\text{tun} +\Gamma_i^\text{rec}$.

We can estimate the tunneling rate $\Gamma_s^\text{tun}$ for a
single particle $s$ (electron or hole) using the tunneling
Hamiltonian.\cite{Bardeen,Luyken99} The probability of 0D to 3D
tunneling for a single particle is\cite{Luyken99}
$
\Gamma_{s}^\text{tun}=\frac{2\pi}{\hbar}\sum_{\alpha}\left|
\langle\Psi_{s}|V_{\text{Dot}}
|\Psi_{\alpha}\rangle\right|^2\delta(E_{s}-E_{\alpha}) \, ,
\label{eq3}
$
where $\Psi_{s}$ and $E_{s}$ is the single particle wave function
(for electron or hole) localized in the QD and its respective
energy, while $\Psi_\alpha$ and $E_\alpha$ are correspondingly in
the contacts. In $\Psi_\alpha$ we use plane waves for the in-plane
motion, and the $z$-component was calculated using the linear
potential from the applied gate voltage and an exponential decay
function in the barrier region.\cite{Landau} $\Psi_{s}$ is
strongly localized in the quantum dot, and it is modelled as
$\Psi_{s}(r,z)=\frac{1}{\sqrt{\pi}l_{s}}\exp({-\frac{r^2}
{2l_0^2}})\chi(z)$, the product of a ground state
harmonic-oscillator function which describes the lateral motion in
a quantum dot, and $\chi(z)$, the wave function of a square well
potential. We obtain
\begin{equation}
\Gamma_{s}^\text{tun}=\left(\frac{4V_{s}}{\pi\hbar}\right)^2\sqrt{\frac{L^2
m_{s}^*}{2|E_{s}|}}
\exp\left(-\frac{4}{3}\frac{\sqrt{2m_{s}^*}}{\hbar e
F}|E_{s}|^\frac{3}{2}\right), \label{eq4}
\end{equation}
where $m_s^*$ is the effective mass of the particle $s$ in the dot,
$L$ and $V_s$ are the width and depth of the corresponding square
well, and $F$ is the electric field provided by the gate voltage.
This is similar to the WKB tunneling but with well-characterized
energy dependent prefactor.

One can consider as a good first approximation that the
electron-hole interaction produces a shift (the exciton binding
energy) in the single particle energy level $E_s$ in the quantum
dot, and use Eq.\ \ref{eq4} to obtain the rates with and without
electron-hole interaction. Note that this estimate of the carrier
tunneling time $\tau_s^\text{tun}=1/\Gamma_s^\text{tun}$ cannot be
directly compared with the experimental linewidth of the
photocurrent peak since that reflects the time for \emph{both}
particles to tunnel out of the system (even if dominated by the
faster rate). Using parameters from Ref.\
[\onlinecite{Findeis01prb}] and the assumption that $V_e\simeq 3V_h$
(where $V_e$ ($V_h$) is the electron (hole) depth of the square well
profile), we obtain that the tunneling times for electron or hole
are similar when there is electron-hole interaction (this of course
depends on QD parameters). After one of the particle tunnels, the
remaining particle tunnels out faster, as it no longer experiences
the electron-hole interaction. The rates we consider here,
$\Gamma_x^\text{tun}$ for the exciton and $\Gamma_b^\text{tun}$ for
the biexciton, are the rates for all particles (electrons and holes)
to leave the system, and they are obtained by solving a separate
density matrix equation including the different tunneling rates for
individual single-particle states. For the exciton state for
instance, we solve a four dimensional density matrix which includes
the vacuum, exciton, electron (when the hole leaves the system
first) and hole states (when the electron leaves the system first).
Solving this density matrix one can evaluate the equivalent rate for
a two level system that reproduces the multi-path process. We obtain
$\Gamma_x^\text{tun}=1/\tau_x^\text{tun}$, where
$\tau_x^\text{tun}\simeq12$ ps is the time for both particles to
leave the system by tunneling, and is consistent with the
photocurrent signal linewidth ($\simeq10$ ps). We can consider this
tunneling as a single process because the charged exciton, which
could be created from a single particle state (after one particle
leaves the QD), is $\simeq4.6$ meV out of resonance from the exciton
state excitation for this kind of QD,\cite{Findeis01prb} and cannot
be created during the pulse duration which is much faster than the
single particle tunneling time. A similar description applies to the
biexciton state. This allows us to write a simple equation for the
photocurrent signal in terms of the exciton and biexciton
contributions as we now describe.

\begin{figure}[tbp]
\includegraphics*[width=1.0\linewidth]{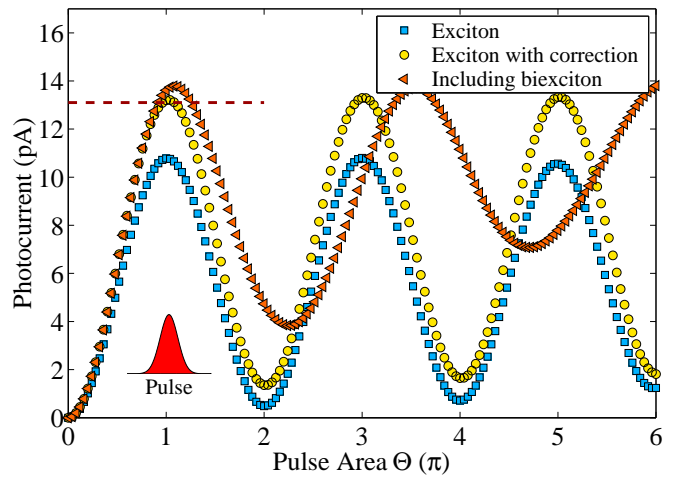}
\caption{Photocurrent signal as function of pulse area $\Theta$
for 1 ps pulse showing different contributions of Eq.\ \ref{eq6}.
Blue squares show results neglecting biexcitons and without the
second integral in Eq.\ \ref{eq6}; yellow circles include the
latter. Triangles show results including biexcitons and all terms
in Eq.\ \ref{eq6}. Dashed line at $I_\text{PC}=fq\simeq13.1$ pA
indicates optimum value for photocurrent if only exciton channel
is considered.} \label{fig3}
\end{figure}

As the photocurrent is a signal induced by the particles that
tunnel out, when the next pulse arrives the system is already in
the vacuum state $|0\rangle$ (meaning that they are different
processes that account for a statistical average), we can write
the expression for the photocurrent as
\begin{equation}
I_\text{PC}=fq\left[\Gamma_b^\text{tun}\int_{-\infty}^\infty\rho_{bb}(t)dt
+\Gamma_x^\text{tun}\int_{-\infty}^\infty\rho_{xx}(t)dt\right],
\label{eq5}
\end{equation}
where $f$ is the repetition frequency of the pulse sequence (we
use $f=82$ MHz as in Zrenner's experiment \cite{Zrenner02}) and
$q$ is the electronic charge. One can also write this as
\begin{eqnarray}
I_\text{PC}&=&fq\Bigr[\Gamma_b^\text{tun}\int_{-\infty}^\tau\rho_{bb}(t)dt
+\Gamma_x^\text{tun}\int_{-\infty}^\tau\rho_{xx}(t)dt \nonumber \\
&&+\frac{\Gamma_b^\text{tun}}{\Gamma_b} \rho_{bb}(\tau) +
\frac{\Gamma_x^\text{tun}}{\Gamma_x}\left(\rho_{xx}(\tau) +
\rho_{bb}(\tau)\right)\Bigr], \label{eq6}
\end{eqnarray}
where $\tau$ is the time at end of the pulse. Notice that the
result is not a simple summation of exciton and biexciton
contributions, but it is rather a mixture. This is expected, as
the biexciton occupation after the pulse eventually decays to the
exciton state, and contributes to the exciton part of the
photocurrent. It is important to point out that even if we ignore
the biexciton contribution to Eq.\ \ref{eq6}, it still exhibits a
term not considered before:\cite{Zrenner02} the contribution
during the pulse given by the second integral in Eq.\ \ref{eq6}.
This contribution can indeed be small depending on the tunneling
time and/or pulse duration. However, it is not the case here, as
shown in Fig.\ \ref{fig3}, where the different Eq.\ \ref{eq6}
contributions to the photocurrent are shown. Blue squares show the
results without the contribution during the pulse (the second
integral in Eq.\ \ref{eq6}) and neglecting the biexciton state in
the entire simulation.  The yellow circles show the same
simulation but including the extra integral term. The difference
between the two traces is noticeable and cannot be ignored.
Including the biexciton in the simulation and using the full Eq.\
\ref{eq6} results in a photocurrent trace that exhibits decay of
the oscillations, but increases slightly with the pulse area,
which is not consistent with experimental observations.

\begin{figure}[tbp]
\includegraphics*[width=1.0\linewidth]{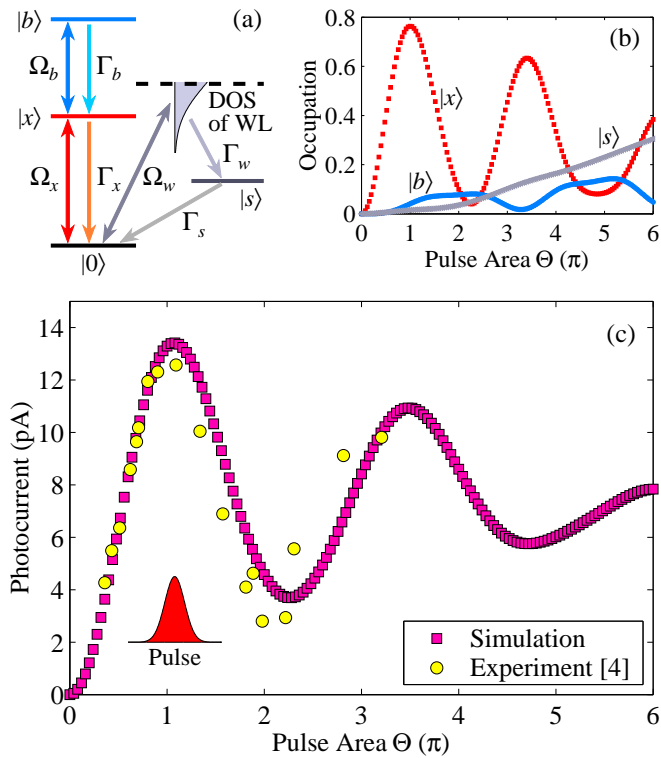}
\caption{(a) Level structure and couplings including nonresonant
excitation to wetting layer WL. (b) Occupation probability of
states considered in model as function of pulse area $\Theta$. (c)
Photocurrent signal vs.\ $\Theta$ for 1 ps pulse. Squares show
results of our model and circles are experimental data from Ref.\
[\onlinecite{Zrenner02}]. Notice this model reproduces
experimental results.} \label{fig4}\end{figure}

Experimental results exhibit a decay of the oscillation with
increasing pulse area, which cannot be obtained from our model so
far. Notice that any kind of additional \textit{constant} decay or
dephasing cannot explain the experimental trace either, since
increasing the pulse area makes this quantity smaller compared with
the pulse intensity, and the Rabi oscillations are enhanced overall.
This behavior indicates the presence of other levels that contribute
to decoherence in the system. \footnote{Note that acoustic phonons
can be another source of decoherence for Rabi oscillations.  See J.
Forstner \textit{et al.}, Phys. Rev. Lett. 91, 127401 (2003); V. M.
Axt, P. Machnikowski, and T. Kuhn, cond-mat/0408322.} It has been
recently demonstrated that the level continuum in the wetting layer
(WL) plays an important role in the background absorption in
self-assembled QDs at high energies.\cite{Vasanelli02} We here
include the dephasing produced by the WL continuum, which in essence
is being populated non-resonantly by the light pulse and then
contributes to the dephasing of all the low-energy excitations. It
is possible to excite electrons from a bound state in the QD
(valence band) to the continuum of the WL (conduction) and from the
continuum (valence) to bound (conduction).\cite{Vasanelli02} We can
model the WL as a unique level detuned from the exciton states by
$\delta_w$, and coherently connected to the ground state by a dipole
interaction $\Omega_w$. This results in an additional term to the
Hamiltonian $H_0$ given by $H_{w}=\delta_{w}|w\rangle \langle
w|-\frac{1}{2}\Omega_w(t)|0\rangle \langle w|+h.c.$. The electron or
hole in this WL state will scatter quickly and leave behind the
other localized single particle $s$. We describe this process by one
decay rate $\Gamma_w$, which connects the WL states to $s$. This
process is shown in Fig.\ \ref{fig4}(a) where it is represented by a
one-direction arrow. Eventually the electron or hole in this $s$
state will tunnel and the system will return to its ground state.
This connection is modelled by $\Gamma_s$, and given by Eq.\
\ref{eq4}. Following the same analysis as for the biexciton, we
would not expect to populate these levels for low intensity, since
they are apparently far detuned from the laser energy. However, the
WL level broadening allows this process to take place, resulting in
these extra levels being populated with increasing pulse intensity,
and a growing background signal in the photocurrent (as that
subtracted in experiments). This is clearly shown in Fig.\
\ref{fig4}(b), where we compare the occupation of some states of the
system, with $\delta_w=20$ meV, $\Gamma_w=40$ meV,
$\Omega_w=\Omega_x$, and, $\Gamma_s=\Gamma_{s}^\text{tun}$ given by
Eq.\ \ref{eq4}. Figure \ref{fig4}(c) shows the resulting
photocurrent obtained from Eq.\ \ref{eq6}, without the background
contribution of the WL. The photocurrent exhibits a decay of the
oscillation induced by the nonresonant excitation to the WL and it
fits quite well the experimental result (although no fine tuning of
parameters has been done).\cite{Zrenner02}

We have described the WL by an additional level that is coherently
excited out of resonance.  In reality, the WL has a broad continuous
distribution of levels resulting in an incoherent pump excitation,
since we have a one-way excitation, as can be seen in Fig.\
\ref{fig4}(a,b) (the population of state $s$ only increases).
Therefore, we can propose an equivalent model that involves a
leakage rate $\Gamma_\text{WL}$ connecting the ground state
$|0\rangle$ to the WL continuum $
\Gamma_\text{WL}=\frac{2\pi}{\hbar}\sum_{\nu}\left|
\frac{1}{2}\langle\nu|\vec{\mu}\cdot \vec{E}(t)
|0\rangle\right|^2\delta(E_{\nu}-E_{0}-\hbar\omega), \label{eq7} $
where the summation is over all $\nu$ levels that compose the WL.\@
This can be written as
\begin{equation}
\Gamma_\text{WL}=\frac{\pi}{2\hbar} \, \rho_\text{WL} \, \mu_w^2
\, E(t)^2 \, , \label{eq8}
\end{equation}
where $\rho_\text{WL}$ is the WL density of states, and $\mu_{w}$ is
the effective dipole moment connecting the WL continuum to the dot
ground state. The master equation (\ref{eq2}) with rate (\ref{eq8})
gives quantitatively similar results to those shown in
Fig.~\ref{fig4}(c) if we use a density of states
$\rho_\text{WL}=\frac{1}{\pi}\frac{\Gamma_w/2}{\Gamma_w^2/4+\delta_w^2}$,
where $\delta_w$ is the WL detuning with the laser, and $\Gamma_w$
is the broadening of the WL levels from our previous model. Notice
that this channel naturally results in an intensity dependent
decoherence, as that described recently.\cite{Wang04,Brandi04}

We have presented a realistic model to describe the Rabi
oscillations observed in a single QD photodiode in the presence of
pulsed light. Using a model that includes multi-exciton and WL
states, a density matrix formalism shows that the two-level model
breaks down and gives an imprecise interpretation of the results.
Full inclusion of the biexciton allowed us to derive an expression
for the photocurrent signal that corrects the simplest model used
in the literature, and more precisely assesses the efficiency of
the level rotation. Our study of the damping mechanisms makes us
conclude that the only possibility to explain the damping with
increasing pulse area (and fixed pulse duration) is an
off-resonant excitation to a different level. This is identified
most likely as the WL continuum of levels. This would also explain
the background in the photocurrent signal observed in
experiments.\cite{Beham03}

This work was partially supported by FAPESP, the US DOE grant no.\
DE--FG02--91ER45334, and the Indiana 21st Century Fund.  We thank
C.J. Villas-B\^{o}as, N. Studart, A. Muller, P. Bianucci, A.H.
MacDonald, and L.E. Oliveira for helpful discussions.

\end{document}